# Fast time resolved techniques as key to the understanding of energy and particle transport in HPPMS-plasmas

C. Maszl, W. Breilmann, L. Berscheid, J. Benedikt and A. von Keudell

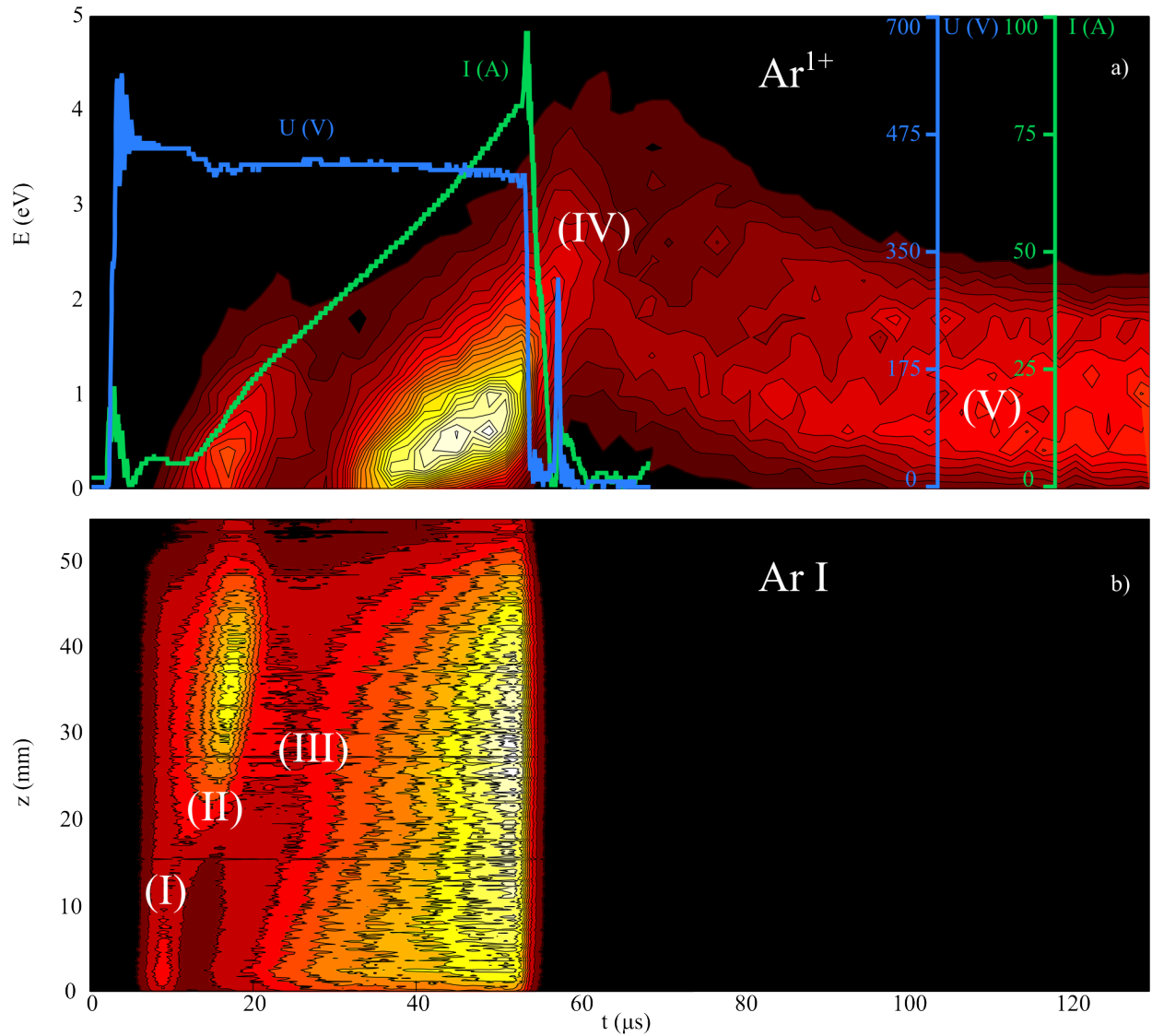

Fig. 1a) Contour plot of the ion energy distribution function (IEDF) of $^{36}Ar^{1+}$. The color depicts the flux in arbitrary units. The discharge current is overlaid in light green and the voltage pulse in light blue. The respective axis can be found on the right-hand side. b) Spatio-temporal diagram for an Ar I line (760 nm). The z-axis starts at the top of the magnetron and ends at the orifice of the EQP. The intensities were extracted at the symmetry axis of the magnetron with a time resolution of 2 µs. The colors are photon counts in arbitrary units. All color scales are linear. The different labels (I)-(V) are explained in the text.

Manuscript received X month 2013; revised X month 2014.
C. Maszl, W. Breilmann, L. Berscheid, J. Benedikt and A. von Keudell are with the Research Department Plasmas with Complex Interactions,
Ruhr-University Bochum, Institute for Experimental Physics II, D-44780 Bochum, Germany
This project is supported by the DFG within the framework of the Coordinated Research Center SFB-TR 87 and the Research Department "Plasmas with Complex Interactions" at Ruhr-University Bochum.
Publisher Identifier S XXXX-XXXXXXX-X

is averaged over 1500 pulses. Fig. 1b) shows a spatio-temporal representation of the Ar I line emission extracted from the central axis of the magnetron for every time step at a resolution of 1 μs. The camera was adjusted to the maximum intensity at the end of the pulse and the data is presented with a linear color scale. At 8 μs, a burst of light becomes visible (I in Fig. 1b). This is identified as a trace of hot electrons from the ignition process. At around 15 μs, a fast upward propagating structure can be observed (II in Fig. 1b). The visible emission pattern is a signature of an ion-acoustic solitary wave in the neutral argon background which travels, depending on the pressure, with a few ~km/s from target to substrate [4]. The arrival of these ions can also be seen in the mass spectrometer in Fig. 1a) at around 18 μs. This prominent feature in the data is used to synchronize the camera and the EQP measurements (otherwise the time of flight of particles in the EQP would have to be taken into account separately). As the current increases also the light intensity increases and the plasma expands in the region between the magnetron and the mass spectrometer (III in Fig. 1b). The latter also detects an increase in the argon ion flux as well as a slight increase in energy [5]. As the voltage is switched off, there is a jump in the energy of the ions (~1.5 eV, IV in Fig. 1a). Fast electrons are lost quickly after plasma shutoff resulting in an increased plasma potential. The remaining ions are therefore accelerated towards the chamber walls. Whereas the afterglow in the ICCD measurements is only visible for a few μs, ions at low energy in the EQP can be observed up to 700 μs. At around 80 μs the ion flux increases again (V in Fig. 1). According to drift-timescales, these slow ions originate from the magnetic trap region where they are bound to the confined electrons via Coulomb forces [5].

Summarizing, EQP and ICCD measurements are valuable diagnostics to gain more insight in the physics of HPPMS plasmas. The identification of the temporal distribution of the arriving fluxes of different species and multiply charged ions will help to understand energy dissipation in film growth and to identify operating regimes, which allow films with superior quality.

---

***Abstract*** – High power pulsed magnetron sputtering (HPPMS) plasmas are pulsed discharges where the plasma composition as well as the fluxes and energies of ions are changing during the pulse. The time resolved energy distribution for $Ar^{1+}$ ions was measured and Phase Resolved Optical Emission Spectroscopy (PROES) for the Ar I line at 760 nm was done to get more insight in the transport properties of the plasma forming noble gas. These measurements were performed during HPPMS of titanium with argon at 0.5 Pa. The peak power density during the 50 μs pulses was 1.8 kW/cm². In this contribution we demonstrate how time resolved mass spectrometry and ICCD cameras can be used to shed more light on energy and particle transport in HPPMS-plasmas.

In HPPMS the plasma is operated only during short pulses of a few μs, followed by long off phases to allow the metal target to cool down. Peak power densities of ~kW/cm² and ionization degrees of the metal vapor of ~80-85% can be achieved. It is believed that the high content of energetic ions is responsible for the resulting excellent thin film properties. However, such pulsing schemes have the disadvantage that the growth rate is reduced compared to direct current magnetron sputtering at same average powers. This is often attributed to the return effect of ions or the production of multiply charged ions [1]. To analyze transport properties of the plasma forming gas to the substrate, we employ time resolved diagnostics like mass spectrometry [2] or the rotating shutter to study the growth rate during a pulse [3].

In this study, a 2-inch magnetron with a titanium target and argon as plasma forming gas were used. The chamber base pressure was $1.5 \times 10^{-4}$ Pa. During plasma operation the pressure was 0.5 Pa at an argon flow rate of 20 sccm. A rectangular voltage pulse was applied for 50 μs with a repetition frequency of 300 Hz (duty cycle 1.5%). The resulting current has a triangular shape with a peak value of 81 A measured before the overshoot at the end (Fig. 1a). This yields a peak power density of 1.8 kW/cm² and an average power of 275 W. Current and voltage were measured directly at the ouput of the SPIK1000A pulse unit. The used VI-probe consists of a LEM LA305-S current transducer with a response time $t_r$<1 μs and a dI/dt>100 A/μs and a LEM CV3-1500 voltage transducer with $t_r$=0.4 μs and dV/dt =900 V/μs. A Hiden (Electrical QuadruPole) EQP 300 HE was used to measure particle fluxes and the ion energy distribution function (IEDF) of $^{36}Ar^{1+}$. The 100 μm orifice of the EQP is mounted in line of sight facing the racetrack in a distance of 8 cm above the target. PROES was done with an ICCD camera and a 760 nm bandpass interference filter (FWHM 10 nm) to study the optical emission of an Ar I line. With this technique it is possible to study spatial light emission during the different phases of HPPMS

Fig. 1a) depicts the applied voltage pulse (light blue), the resulting current (light green) and a contour plot with a linear color scale of the IEDF measured with a temporal resolution of 2 μs. The IEDF for each time step